\documentclass[%
 reprint,
 amsmath,amssymb,
aps,
prb,
]{revtex4-1}

\usepackage{graphicx}
\usepackage{dcolumn}
\usepackage{bm}

\begin{document}

\title{Plasmonic enhancement of spatial dispersion effects in prism coupler experiments}

\author{Armel Pitelet}
\author{Emilien Mallet}
\author{Rabih Ajib}
\author{Caroline Lema\^itre}
\author{Emmanuel Centeno}
\author{Antoine Moreau}
\email{antoine.moreau@uca.fr}
\affiliation{%
 Universit\'e Clermont Auvergne, CNRS, Institut Pascal, 63000 Clermont-Ferrand, France
}%
\date{\today}

\begin{abstract}
Recent experiments with film-coupled nanoparticles suggest that the impact of spatial dispersion is enhanced in plasmonic structures where high wavevector guided modes are excited. More advanced descriptions of the optical response of metals than Drude's are thus probably necessary in plasmonics. We show that even in classical prism coupler experiments, the plasmonic enhancement of spatial dispersion can be leveraged to make such experiments two orders of magnitude more sensitive. The realistic multilayered structures involved rely on layers that are thick enough to rule our any other phenomenon as the spill-out. Optical evanescent excitation of plasmonic waveguides using prism couplers thus constitutes an ideal platform to study spatial dispersion. 
\end{abstract}

\maketitle

Spatial dispersion, {\em i.e.} the dependency of the permittivity on the wavevector, arises in metals because of the repulsive interaction between the free electrons inside the jellium. This phenomenon puts a limit to the validity of Drude's model\cite{drude1900elektronentheorie}, which has proven extremely accurate in plasmonics for more than a century now. This success can be related to the fact that Drude's model is the zero-th order term of all the more advanced descriptions of  the free electron gas\cite{fuchs,feibelman1975microscopic,eguiluz1976hydrodynamic,boardman82,crouseilles08}. The first-order correction to the model, however, is intrinsically nonlocal: because of the repulsion between free electrons, the metal can still be described using an effective polarization, but the polarization in a given point depends on the electric field in the surroundings - at a typical distance close to the free mean path of the electrons.

The idea that spatial dispersion could have an impact on the optical response of metals and on surface plasmons dates back to the 1960s. Very advanced models\cite{feibelman1975microscopic,wang2011foundations}, including the hydrodynamic model\cite{fuchs,eguiluz1976hydrodynamic}, have been proposed at the time to tackle the problem. However, it became clear in the 1980s that no optical experiment with noble metals could show an impact of spatial dispersion large enough to threaten the predominance of Drude's model in plasmonics with an optical excitation. While spatial dispersion and more advanced models for surface plasmons\cite{liebsch1993surface,liebsch1993surface1,CHANG2003353} continued to be thoroughly studied in surface science especially using EELS\cite{rocca1995low,rocca1995surface,park2009plasmon}, the consensus seemed to be that Drude's model is largely sufficient in plasmonics\cite{boardman82,forstmann86}.

For many in this community, the experiments showing Drude's model failure with large enough nanoparticles excited optically\cite{ciraci2012probing,ciraci2014film} thus came as a surprise. These experiments renewed the interest paid to the hydrodynamic model\cite{ciraci2013hydrodynamic}, which proved accurate enough, and to the enhancement of spatial dispersion brought by small gaps between metals instead of relying on tiny particles\cite{scholl2012quantum,raza2013nonlocal}. However, the sub-nanometer gaps required to observe the impact of spatial dispersion raised some skepticism\cite{hajilsalem14}, all the more so that in that case the effects studied in surface science, like the spill-out, are likely to intervene strongly\cite{esteban2012bridging,savage2012revealing,teperik2013robust}. A lot of work has subsequently been devoted to (i) study theoretically in which situations spatial dispersion is likely to have an impact\cite{fernandez2010collection,wiener2012nonlocal,moreau13,dechaux16} (ii) develop numerical tools based on the hydrodynamical model\cite{toscano2012modified,benedicto2015numerical,Schmitt2016396} to accurately predict these effects in complex geometries and finally (iii) to better understand the fundamental reasons why the hydrodynamic model, while it presents well documented deficiencies\cite{haberland2013looking}, seems accurate enough for noble metals in plasmonics\cite{toscano2015resonance,ciraci2016quantum}.

Here we propose to use the classical prism-coupler configuration\cite{lopezrios,CHANG2003353} to excite high wavevector plasmonic modes, in order to enhance the effects of spatial dispersion alone. We predict, relying on the hydrodynamic model and accurate material parameters, that the impact of spatial dispersion will be two orders of magnitude larger than what can be reached with simple surface plasmons. The plasmonic resonances excited can be linked to the excitation of Gap-Plasmons\cite{bozhevolnyi2007general,moreau13} (GP), Long-Range Surface Plasmons (LRSP) and Short-Range Surface Plasmons\cite{berini2009long,PhysRevB.61.10484,raza2013nonlocal,pitelet17}(SRSP) supported by multilayered structures with dimensions that are large enough to exclude other phenomenon like the spill-out\cite{teperik2013robust}. Were such experiments to be conducted, they should allow to measure accurately crucial parameters of the hydrodynamic model.

\begin{figure}[!h]
\includegraphics[width=\linewidth]{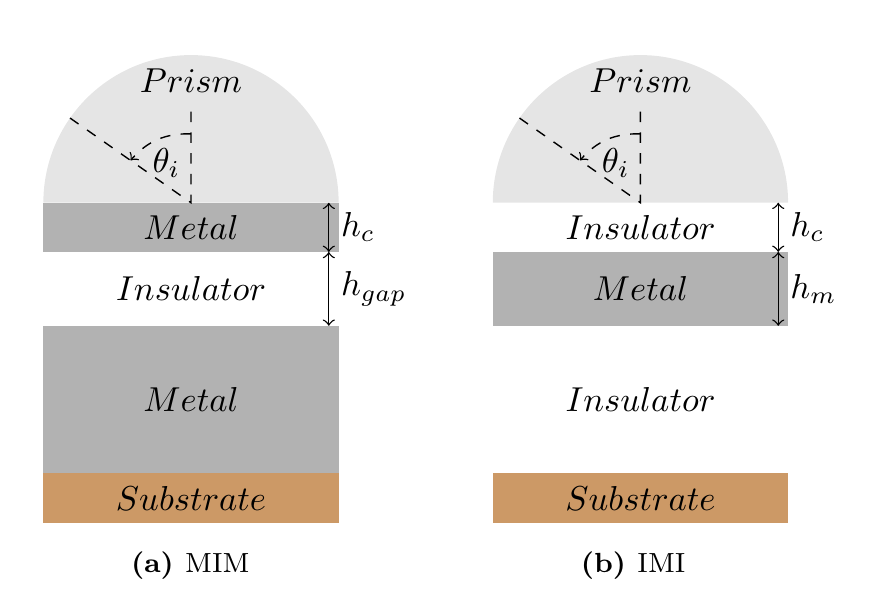}
\caption{Schematic representation of the prism couplers: \textbf{(a)} resembles the Otto configuration, i.e a prism on top of a metallic slab which ensures coupling to the gap-plasmon. while\textbf{(b)} resembles the  Kretschmann-Raether configuration, i.e a prism on top of  a dielectric layer through which the guided modes of an IMI structure are coupled.}\label{schemas}
\end{figure}

Fig.~\ref{schemas} shows the prism-coupler structures for which the coupling condition can be written as a  relation between the effective index $n_{eff}=k_{x}/k_0$ of a guided mode (characterized by a wavevector $k_{x}$  at a frequency $\omega = k_0 c$) to the prism index $n_p$ and angle of incidence at the prism bottom interface $\theta_i$ 
\begin{align}
	n_{eff} = n_p sin(\theta_i)\label{snell}
\end{align}
showing it is a way to measure the wavevector directly. Any change in $k_{x}$ induced by spatial dispersion leads to a discrepancy between the coupling angle predicted using Drude's model and taking nonlocality into account. We underline however, as will be clear in the following, that the change in the coupling angle is not the only change brought by nonlocality.

In the hydrodynamic model\cite{boardman82, moreau13,ciraci2013hydrodynamic,ciraci2016quantum,toscano2015resonance}, the link the electric field $\textbf{E}$ in the metal to the induced electronic current $\textbf{J}$ is written as
\begin{align}
   -\beta^2 \nabla(\nabla.\mathbf{J})+\mathbf{\ddot{J}}+ \gamma \mathbf{\dot{J}}=\epsilon_0 \omega_p^2 \mathbf{\dot{E}} \label{eq_hydro}
\end{align}
$\omega_p$ being the plasma frequency, $\epsilon_0$ the vacuum permittivity and $\beta$ the non-local parameter which describe the interactions between free electrons in metal. A special attention must be paid to (i) the material parameters  -- since our version of the hydrodynamic model makes an explicit distinction between the response of free electrons, considered as nonlocal, and the response of core electrons, considered as purely local -- and 
(ii) the necessary additional boundary condition -- here we impose that no electron is allowed to leave the metal. For the material parameters, we rely on a Brendel-Bormann model that has been proven to be particularly accurate\cite{rakic98}. Finally, regarding parameters, several theoretical expressions exist for the nonlocal parameter $\beta$. We use a value of $\beta = 1.35\times 10^{6}~ m.s^{-1}$ coming from the best available experimental data\cite{ciraci2014film}. We use this value to assess whether nonlocality can have an impact on the optical response, but we stress that the setups we proposes actually constitute a way to measure this crucial parameter.

The reflectance of the structures is computed using Moosh, an open code\cite{defrance16} that is able to take spatial dispersion into account in the framework of the hydrodynamic model through the use of a specifically designed scattering matrix algorithm\cite{benedicto2015numerical}. The situations described below result from  a choice of the geometrical parameters which maximizes the impact of spatial dispersion -- but the phenomenon is always easy to spot. 

We first study the optical excitation of a Gap-Plasmon resonance in a gap formed by an insulator layer sandwiched between two metallic slabs (see Fig. \ref{schemas} (a), MIM structure). For the GP propagating in a dielectric layer with a permittivity $\epsilon_d$ between two metals with a permittivity 
$\epsilon_m = 1 + \chi_f+ \chi_b$ where $\chi_f$ and $\chi_b$ are the susceptibilities linked to the free and bound electrons respectively, the dispersion relation for a mode at frequency $\omega$ and with a wavevector $k_x$ can be written\cite{moreau13,raza2013nonlocal}
\begin{equation}
\frac{\kappa_d}{\epsilon_d} \tanh \frac{\kappa_d h}{2} + \frac{\kappa_m}{\epsilon_m} = \Omega
\end{equation}
where $\kappa_i = \sqrt{k_x^2-\epsilon_i k_0^2}$ with $i=d,m$, $k_0=\omega/c$ and $\Omega = \frac{k_x^2}{\kappa_l} \left(\frac{1}{\epsilon_m}-\frac{1}{1+\chi_b}\right)$ with $\kappa_l^2=k_x^2 + \frac{\omega_p^2}{\beta^2}\left(\frac{1}{\chi_f}+\frac{1}{\chi_b}\right)$. The parameter $\Omega$ vanishes when there is no spatial dispersion, but increases with $k_x$, which leads to expect a larger impact for higher wavevector. The wavevector increases when $h$ decreases, so that decreasing $h$ leads to a higher discrepancy between Drude's model and the hydrodynamic model. 

\begin{figure}
\includegraphics[width=\linewidth]{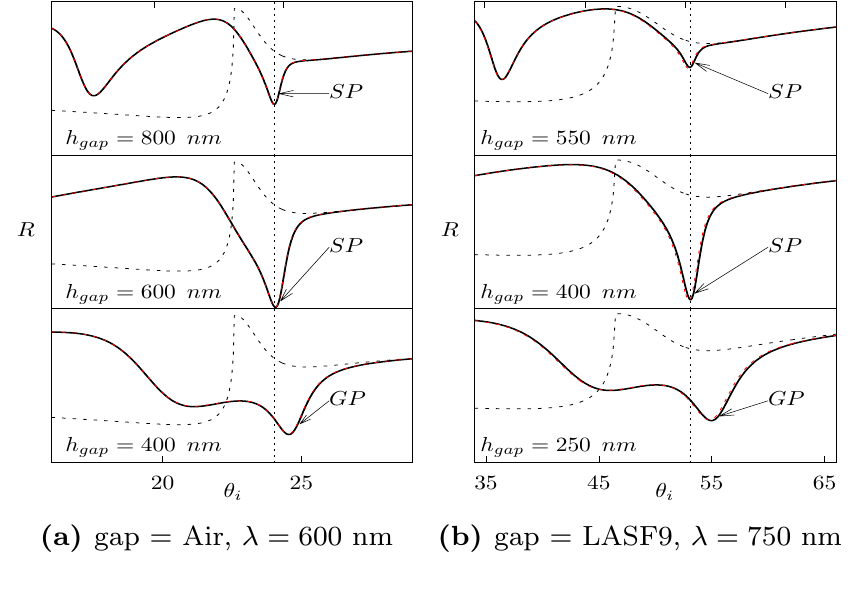}
\caption{\textbf{(a)} : Reflectivity in the local approximation (solid black line) and with the hydrodynamic model (dashed red line) for the MIM structure with air. The dashed black line corresponds to the coupler without the second metallic interface. The vertical dotted line shows where the undisturbed surface plasmon is expected. For large gaps only the surface plasmon on the lower interface is excited. For the thinner gap the symmetric and antisymmetric modes can clearly be seen. The angle of excitation for the symmetric mode is higher than for the SP signaling the gap-plasmon regime. (b) Same situation with a dielectric loaded gap and a lower frequency. The impact of nonlocality begins to be noticeable because the resonance is at larger incidence angle.
\label{coupling}}
\end{figure}

The effective index of the GP is controlled by the thickness of the dielectric $h_{gap}$ supporting it\cite{bozhevolnyi2007general}. As $h_{gap}$  decreases, the wavevector of the GP increases and even diverges when $h_{gap}\rightarrow 0$. For an ultra-thin $h_{gap}$, the GP wavevector is large enough to make the GP sensitive to nonlocality and thus to deviate from Drude's model prediction regarding the coupling angle\cite{moreau13}. Defining the coupling angle as the angle for which reflectivity reaches its lowest value, we define $\Delta \theta_i$ as the difference between the local and nonlocal coupling angle - the nonlocal angle being always the smallest.
In order to reach a meaningful difference we have to push the GP wavevector as high as possible by using the smallest dielectric thickness. We are however limited by the prism refractive index which determines the maximum reachable effective index. That is the reason why we consider $TiO_2$\cite{DeVore:51} prisms: they present the highest possible refractive index in the visible range and are commercially available.

We use $Au$ as metal and air as dielectric (a setup for which the gap can thus be changed progressively). Fig. \ref{coupling} shows how, when the gap is decreased, the two interfaces become coupled, and the entrance in the gap-plasmon regime when the effective index of the symmetric mode, and thus the angle of excitation, increase clearly compared to the surface plasmons. We stress that, when the interfaces are decoupled, the discrepancy between the local and the nonlocal descriptions is below $0.01^\circ$, a result coherent with previous studies of the phenomenon\cite{CHANG2003353}.

\begin{figure}[t]

\includegraphics{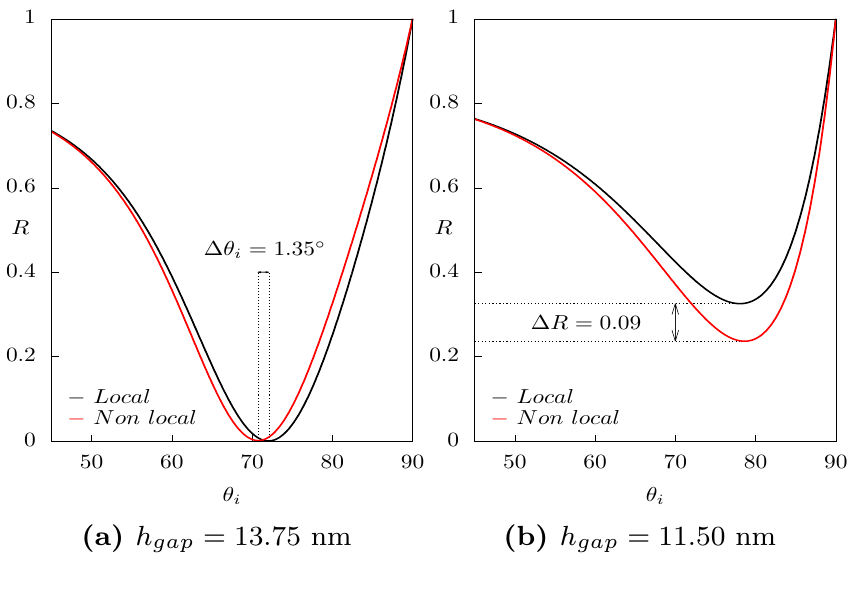}\label{Air1375}

\caption{Reflectivity of the structure shown on Fig.~\ref{schemas}(a) as a function of the incident angle $\theta_i$ for two value of $h_{gap}$. The wavelength of the incident light is $\lambda = 600$ nm. The prism is made of $TiO_2$ with a permittivity $\epsilon_p = 6.78$, the metal is $Au$ with a permittivity $\epsilon_m = -8.44+1.41i$ and the dielectric is air. The upper metallic layer thickness is $h_c=18$ nm. The resonance corresponds to the excitation of a gap-plasmon.}\label{mimAir}
\end{figure}

Fig~\ref{mimAir} the reflectivity of our structure as a function of the incident angle for a wavelength $\lambda$ of 600 nm. With $h_{gap} = 13.75$ nm ( see left part of Fig~\ref{mimAir}), non-locality causes a shift of the coupling angle of more than $1^{\circ}$, which is a two orders of magnitude increase compared to the surface plasmon. Here, the position of the reflectivity minimum is close to the coupling angle that can be determined from the dispersion relation -- a regime that we thus call the plain coupling regime and for which the effective index of the mode is lower than the prism index.

%
%
%
%

For a slightly thinner gap  ($h_{gap} = 11.50$ nm), the GP wavevector is beyond the maximum incident wavevector reachable but near enough to still impact the reflectivity. The reflectivity still presents a minimum, but its position cannot be predicted using the effective index of the mode any more. The presence of the guided mode, even if it can be only imperfectly coupled is still responsible for the minimum. We attribute the difference between Drude's model prediction and the predictions of the hydrodynamic model to the fact that the effective index of the gap-plasmon is always lower when spatial dispersion is taken into account. The resonance being further off with a local description, the minimum is higher. While the presence of a closer resonance in the nonlocal case makes the minimum lower. This situation, that we call the near coupling regime, gives rise to a discrepancy $\Delta R$ for the minimum of the reflectivity, shown Fig~\ref{mimAir}. Such a difference appears at grazing incidence, but it may eventually be easier to measure than the angular shift in the plain coupling regime. 

For both regimes, the only condition we have to satisfy to study non-locality is to reach a high enough wavevector for the gap-plasmon. We have done this  so far by using an extremely thin gap filled with air, but we now show it is possible to reach similar sensitivity to non-locality relying on much larger thicknesses provided the refractive index of the dielectric is significantly higher than 1. 

\begin{figure}[!h]
\includegraphics{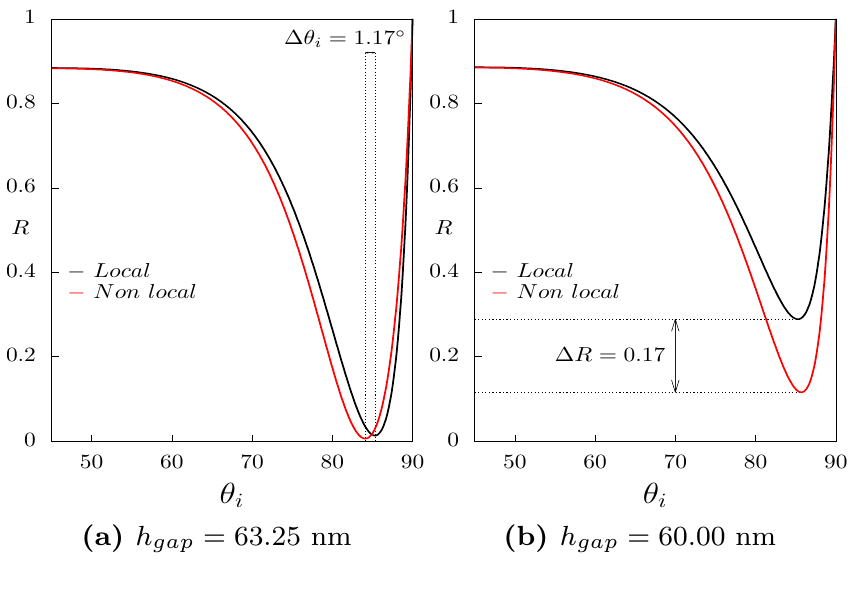}\label{diel6000}
\caption{Reflectivity of the structure shown on Fig. \ref{schemas}(a) as a function of the incident angle $\theta_i$ for two values of $h_{gap}$. The wavelength of the incident light is $\lambda = 750$ nm. The prism is made of $TiO_2$ with a permittivity $\epsilon_p = 6.41$, the metal is $Au$ with a permittivity $\epsilon_m = -18.50+1.50i$ and the dielectric is LASF9 with a permittivity $\epsilon_d = 3.37+1.10e^{-07}i$. The thickness of the upper metallic layer is $h_c=18$ nm.}\label{mimdiel}
\end{figure}

For a fixed wavelength, using such materials actually leads plasmonic guided modes to present higher wavevectors than using air as dielectric for the same gap width. Very similar results to the above ones are obtained for much large thicknesses of more than 60 nm. We have studied the case of a dielectric waveguide filled with LASF9 at a wavelength of  $\lambda = 750$ nm. Exactly as previously, the impact of spatial dispersion is observed in the plain coupling regime, producing a shift of the coupling angle (see Fig.\ref{mimdiel}(a)), and in the near coupling regime (see Fig. \ref{mimdiel}(b)), producing a very large discrepancy in the reflectivity  of $0.17$. This setup presents several advantages as (i) it is probably easier to control the gap over macroscopic distances horizontally if it is filled with a dielectric (ii) using a dielectric allows to reach higher effective index for quite large wavelengths (iii) such a large thickness allows to completely neglect other phenomena like the spill-out, which intervene when extremely small gaps are involved.

\begin{figure}
\centering
\includegraphics[width=8.6cm, height=6cm]{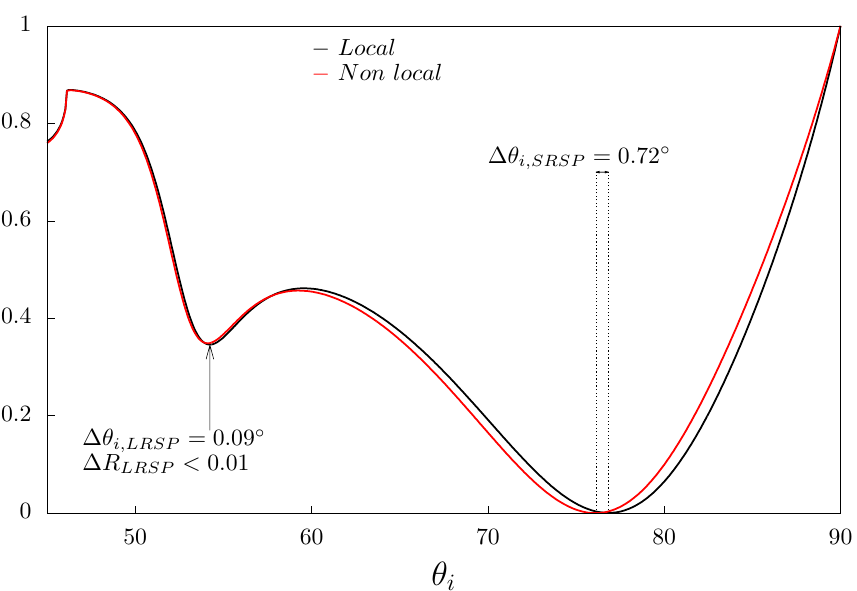}
\caption{Reflectivity of the structure shown on Fig.~\ref{schemas}(b) as a function of the incident angle $\theta_i$ for $h_{m} = 44$ nm. The wavelength of the incident light is $\lambda = 700$. Prism is $TiO_2$ with permittivity $\epsilon_p = 6.50$ metal is $Au$ with permittivity $\epsilon_m = -15.04+1.31i$ and dielectric is LASF9 with permittivity $\epsilon_d = 3.28+1.15\times 10^{-07}i$ (a permittivity very close to the one of other high index dielectrics like $Al_2 O_3$). The thickness of the layer under the prism is $h_c=65$ nm. The two dips are due to the excitation of LRSP (left) in the plain regime and SRSP (right) in the near coupling regime.}\label{imi}
\end{figure}

Finally, we show that it is possible to excite the guided modes supported by the complementary structure: a metallic slab\cite{raza2013nonlocal,pitelet17} buried in a dielectric, a structure we will call IMI shown on Fig. ~\ref{schemas}(b). We underline that the modes of a single metallic slab have been envisaged a long time ago in the framework of EELS\cite{ritchie1962optical}. Compared to the MIM structure, the IMI is able to support not only one but two guided modes, the LRSP and the SRSP. These two modes have two distinct wavevectors which behave differently when the thickness of the metallic slab supporting them varies. The dispersion relation for the symmetric SRSP mode can be written 
\begin{equation}
\frac{\kappa_m}{\epsilon_m} \tanh \frac{\kappa_m h}{2} + \frac{\kappa_d}{\epsilon_d} = \Omega
\end{equation}
and for the antisymmetric LRSP 
\begin{equation}
\frac{\kappa_m}{\epsilon_m} \coth \frac{\kappa_m h}{2} + \frac{\kappa_d}{\epsilon_d} = \Omega.
\end{equation}

As a consequence, while the SRSP tends to behave as the GP, i.e to present a diverging wavevector when $h_{m}$ tends to zero, the LRSP has a lower wavevector which stays much more stable when  $h_{m}$ decreases. The reflectivity of our IMI structure excited through the prism thus presents two dips for two different angles corresponding respectively to the LRSP and SRSP modes. Here too the parameter $\Omega$ appears in the dispersion relation. Since it is larger when the wavevector $k_x$ is larger, the impact of spatial dispersion on the can be expected to be much lower for the LRSP than for the SRSP\cite{raza2013nonlocal,pitelet17}. Fig~\ref{imi} shows that while spatial dispersion has a small impact on the LRSP, there is a clear shift of  $0.72^\circ$ for the higher effective index SRSP when nonlocality in taken into account. We underline that for such a setup the LRSP can thus be used to retrieve the material and geometrical parameters while the SRSP allows for a measurement of $\beta$. 

In conclusion, we have shown how feasible  multilayered structures could be used to enhance the impact of spatial dispersion in metals, through the excitation of three of the most emblematic guided mode in plasmonics: the gap-plasmon, and the short- and long-range surface plasmons. Physically, this enhancement by two orders of magnitude can be directly linked to the high wavevectors these plasmonic guided modes present. It is obtained here for dielectric gaps whose typical thicknesses, ranging from 10 nm to 70 nm are one to two orders of magnitude larger than in  previous experiments\cite{ciraci2012probing,ciraci2014film}. This allows to rule out any role of more complex phenomena like the spill-out of the electron gas outside of the  metal\cite{esteban2012bridging,savage2012revealing,teperik2013robust,ciraci2016quantum} because of the finite extraction work. 

The geometrical parameters for which so strong nonlocal effects can be found suggest 
spatial dispersion has probably to be taken into account in plasmonics for much larger structures than previously thought. We underline that the present setups do not involve a large number of chemically synthesized nanoparticles\cite{scholl2012quantum,ciraci2012probing,ciraci2014film,raza2015nonlocal} and, furthermore, that the working wavelengths are in the red part of the optical spectrum -- which shows that these effects manifest themselves even far below the plasma frequency of metals. 

We hope our work will pave the way for well controlled optical experiments allowing to assess whether or not the hydrodynamic model is an accurate replacement for Drude's in plasmonics. Given the increasing number of devices relying on gap-plasmon excitation in dielectric-filled gaps\cite{akselrod2014probing,lassiter2014third,haffner2015all,nielsen2017giant} such an evolution may be a necessity soon. 

\section*{Funding Information}

This work has been supported by the French National Research Agency, Project “Physics of Gap-Plasmons”  No.
ANR-13-JS10-0003.

\bibliography{sample}

\begin{thebibliography}{47}%
\makeatletter
\providecommand \@ifxundefined [1]{%
 \@ifx{#1\undefined}
}%
\providecommand \@ifnum [1]{%
 \ifnum #1\expandafter \@firstoftwo
 \else \expandafter \@secondoftwo
 \fi
}%
\providecommand \@ifx [1]{%
 \ifx #1\expandafter \@firstoftwo
 \else \expandafter \@secondoftwo
 \fi
}%
\providecommand \natexlab [1]{#1}%
\providecommand \enquote  [1]{``#1''}%
\providecommand \bibnamefont  [1]{#1}%
\providecommand \bibfnamefont [1]{#1}%
\providecommand \citenamefont [1]{#1}%
\providecommand \href@noop [0]{\@secondoftwo}%
\providecommand \href [0]{\begingroup \@sanitize@url \@href}%
\providecommand \@href[1]{\@@startlink{#1}\@@href}%
\providecommand \@@href[1]{\endgroup#1\@@endlink}%
\providecommand \@sanitize@url [0]{\catcode `\\12\catcode `\$12\catcode
  `\&12\catcode `\#12\catcode `\^12\catcode `\_12\catcode `\%12\relax}%
\providecommand \@@startlink[1]{}%
\providecommand \@@endlink[0]{}%
\providecommand \url  [0]{\begingroup\@sanitize@url \@url }%
\providecommand \@url [1]{\endgroup\@href {#1}{\urlprefix }}%
\providecommand \urlprefix  [0]{URL }%
\providecommand \Eprint [0]{\href }%
\providecommand \doibase [0]{http://dx.doi.org/}%
\providecommand \selectlanguage [0]{\@gobble}%
\providecommand \bibinfo  [0]{\@secondoftwo}%
\providecommand \bibfield  [0]{\@secondoftwo}%
\providecommand \translation [1]{[#1]}%
\providecommand \BibitemOpen [0]{}%
\providecommand \bibitemStop [0]{}%
\providecommand \bibitemNoStop [0]{.\EOS\space}%
\providecommand \EOS [0]{\spacefactor3000\relax}%
\providecommand \BibitemShut  [1]{\csname bibitem#1\endcsname}%
\let\auto@bib@innerbib\@empty
\bibitem [{\citenamefont {Drude}(1900)}]{drude1900elektronentheorie}%
  \BibitemOpen
  \bibfield  {author} {\bibinfo {author} {\bibfnamefont {P.}~\bibnamefont
  {Drude}},\ }\href@noop {} {\bibfield  {journal} {\bibinfo  {journal} {Annalen
  der Physik}\ }\textbf {\bibinfo {volume} {306}},\ \bibinfo {pages} {566}
  (\bibinfo {year} {1900})}\BibitemShut {NoStop}%
\bibitem [{\citenamefont {Fuchs}\ and\ \citenamefont {Kliewer}(1971)}]{fuchs}%
  \BibitemOpen
  \bibfield  {author} {\bibinfo {author} {\bibfnamefont {R.}~\bibnamefont
  {Fuchs}}\ and\ \bibinfo {author} {\bibfnamefont {K.~L.}\ \bibnamefont
  {Kliewer}},\ }\href {\doibase 10.1103/PhysRevB.3.2270} {\bibfield  {journal}
  {\bibinfo  {journal} {Phys. Rev. B}\ }\textbf {\bibinfo {volume} {3}},\
  \bibinfo {pages} {2270} (\bibinfo {year} {1971})}\BibitemShut {NoStop}%
\bibitem [{\citenamefont {Feibelman}(1975)}]{feibelman1975microscopic}%
  \BibitemOpen
  \bibfield  {author} {\bibinfo {author} {\bibfnamefont {P.~J.}\ \bibnamefont
  {Feibelman}},\ }\href@noop {} {\bibfield  {journal} {\bibinfo  {journal}
  {Physical Review B}\ }\textbf {\bibinfo {volume} {12}},\ \bibinfo {pages}
  {1319} (\bibinfo {year} {1975})}\BibitemShut {NoStop}%
\bibitem [{\citenamefont {Eguiluz}\ and\ \citenamefont
  {Quinn}(1976)}]{eguiluz1976hydrodynamic}%
  \BibitemOpen
  \bibfield  {author} {\bibinfo {author} {\bibfnamefont {A.}~\bibnamefont
  {Eguiluz}}\ and\ \bibinfo {author} {\bibfnamefont {J.}~\bibnamefont
  {Quinn}},\ }\href@noop {} {\bibfield  {journal} {\bibinfo  {journal}
  {Physical Review B}\ }\textbf {\bibinfo {volume} {14}},\ \bibinfo {pages}
  {1347} (\bibinfo {year} {1976})}\BibitemShut {NoStop}%
\bibitem [{\citenamefont {Boardman}(1982)}]{boardman82}%
  \BibitemOpen
  \bibfield  {author} {\bibinfo {author} {\bibfnamefont {A.~D.}\ \bibnamefont
  {Boardman}},\ }\href@noop {} {\emph {\bibinfo {title} {Electromagnetic
  surface modes}}}\ (\bibinfo  {publisher} {Wiley},\ \bibinfo {year}
  {1982})\BibitemShut {NoStop}%
\bibitem [{\citenamefont {Crouseilles}\ \emph {et~al.}(2008)\citenamefont
  {Crouseilles}, \citenamefont {Hervieux},\ and\ \citenamefont
  {Manfredi}}]{crouseilles08}%
  \BibitemOpen
  \bibfield  {author} {\bibinfo {author} {\bibfnamefont {N.}~\bibnamefont
  {Crouseilles}}, \bibinfo {author} {\bibfnamefont {P.~A.}\ \bibnamefont
  {Hervieux}}, \ and\ \bibinfo {author} {\bibfnamefont {G.}~\bibnamefont
  {Manfredi}},\ }\href@noop {} {\bibfield  {journal} {\bibinfo  {journal}
  {Phys. Rev. B}\ }\textbf {\bibinfo {volume} {78}},\ \bibinfo {pages} {155412}
  (\bibinfo {year} {2008})}\BibitemShut {NoStop}%
\bibitem [{\citenamefont {Wang}\ \emph {et~al.}(2011)\citenamefont {Wang},
  \citenamefont {Plummer},\ and\ \citenamefont {Kempa}}]{wang2011foundations}%
  \BibitemOpen
  \bibfield  {author} {\bibinfo {author} {\bibfnamefont {Y.}~\bibnamefont
  {Wang}}, \bibinfo {author} {\bibfnamefont {E.}~\bibnamefont {Plummer}}, \
  and\ \bibinfo {author} {\bibfnamefont {K.}~\bibnamefont {Kempa}},\
  }\href@noop {} {\bibfield  {journal} {\bibinfo  {journal} {Advances in
  Physics}\ }\textbf {\bibinfo {volume} {60}},\ \bibinfo {pages} {799}
  (\bibinfo {year} {2011})}\BibitemShut {NoStop}%
\bibitem [{\citenamefont {Liebsch}(1993{\natexlab{a}})}]{liebsch1993surface}%
  \BibitemOpen
  \bibfield  {author} {\bibinfo {author} {\bibfnamefont {A.}~\bibnamefont
  {Liebsch}},\ }\href@noop {} {\bibfield  {journal} {\bibinfo  {journal}
  {Physical Review B}\ }\textbf {\bibinfo {volume} {48}},\ \bibinfo {pages}
  {11317} (\bibinfo {year} {1993}{\natexlab{a}})}\BibitemShut {NoStop}%
\bibitem [{\citenamefont {Liebsch}(1993{\natexlab{b}})}]{liebsch1993surface1}%
  \BibitemOpen
  \bibfield  {author} {\bibinfo {author} {\bibfnamefont {A.}~\bibnamefont
  {Liebsch}},\ }\href@noop {} {\bibfield  {journal} {\bibinfo  {journal}
  {Physical review letters}\ }\textbf {\bibinfo {volume} {71}},\ \bibinfo
  {pages} {145} (\bibinfo {year} {1993}{\natexlab{b}})}\BibitemShut {NoStop}%
\bibitem [{\citenamefont {Chang}\ \emph {et~al.}(2003)\citenamefont {Chang},
  \citenamefont {Chiang}, \citenamefont {Leung},\ and\ \citenamefont
  {Tse}}]{CHANG2003353}%
  \BibitemOpen
  \bibfield  {author} {\bibinfo {author} {\bibfnamefont {R.}~\bibnamefont
  {Chang}}, \bibinfo {author} {\bibfnamefont {H.-P.}\ \bibnamefont {Chiang}},
  \bibinfo {author} {\bibfnamefont {P.}~\bibnamefont {Leung}}, \ and\ \bibinfo
  {author} {\bibfnamefont {W.}~\bibnamefont {Tse}},\ }\href {\doibase
  https://doi.org/10.1016/j.optcom.2003.07.048} {\bibfield  {journal} {\bibinfo
   {journal} {Optics Communications}\ }\textbf {\bibinfo {volume} {225}},\
  \bibinfo {pages} {353 } (\bibinfo {year} {2003})}\BibitemShut {NoStop}%
\bibitem [{\citenamefont {Rocca}(1995)}]{rocca1995low}%
  \BibitemOpen
  \bibfield  {author} {\bibinfo {author} {\bibfnamefont {M.}~\bibnamefont
  {Rocca}},\ }\href@noop {} {\bibfield  {journal} {\bibinfo  {journal} {Surface
  science reports}\ }\textbf {\bibinfo {volume} {22}},\ \bibinfo {pages} {1}
  (\bibinfo {year} {1995})}\BibitemShut {NoStop}%
\bibitem [{\citenamefont {Rocca}\ \emph {et~al.}(1995)\citenamefont {Rocca},
  \citenamefont {Yibing}, \citenamefont {de~Mongeot},\ and\ \citenamefont
  {Valbusa}}]{rocca1995surface}%
  \BibitemOpen
  \bibfield  {author} {\bibinfo {author} {\bibfnamefont {M.}~\bibnamefont
  {Rocca}}, \bibinfo {author} {\bibfnamefont {L.}~\bibnamefont {Yibing}},
  \bibinfo {author} {\bibfnamefont {F.~B.}\ \bibnamefont {de~Mongeot}}, \ and\
  \bibinfo {author} {\bibfnamefont {U.}~\bibnamefont {Valbusa}},\ }\href@noop
  {} {\bibfield  {journal} {\bibinfo  {journal} {Physical Review B}\ }\textbf
  {\bibinfo {volume} {52}},\ \bibinfo {pages} {14947} (\bibinfo {year}
  {1995})}\BibitemShut {NoStop}%
\bibitem [{\citenamefont {Park}\ and\ \citenamefont
  {Palmer}(2009)}]{park2009plasmon}%
  \BibitemOpen
  \bibfield  {author} {\bibinfo {author} {\bibfnamefont {S.~J.}\ \bibnamefont
  {Park}}\ and\ \bibinfo {author} {\bibfnamefont {R.~E.}\ \bibnamefont
  {Palmer}},\ }\href@noop {} {\bibfield  {journal} {\bibinfo  {journal}
  {Physical review letters}\ }\textbf {\bibinfo {volume} {102}},\ \bibinfo
  {pages} {216805} (\bibinfo {year} {2009})}\BibitemShut {NoStop}%
\bibitem [{\citenamefont {Frostmann}\ and\ \citenamefont
  {Gerhardts}(1986)}]{forstmann86}%
  \BibitemOpen
  \bibfield  {author} {\bibinfo {author} {\bibfnamefont {F.}~\bibnamefont
  {Frostmann}}\ and\ \bibinfo {author} {\bibfnamefont {R.~R.}\ \bibnamefont
  {Gerhardts}},\ }\href@noop {} {\emph {\bibinfo {title} {Metal optics near the
  plasma frequency}}},\ Vol.\ \bibinfo {volume} {109}\ (\bibinfo  {publisher}
  {Springer-Verlag},\ \bibinfo {year} {1986})\BibitemShut {NoStop}%
\bibitem [{\citenamefont {Cirac{\`\i}}\ \emph {et~al.}(2012)\citenamefont
  {Cirac{\`\i}}, \citenamefont {Hill}, \citenamefont {Mock}, \citenamefont
  {Urzhumov}, \citenamefont {Fern{\'a}ndez-Dom{\'\i}nguez}, \citenamefont
  {Maier}, \citenamefont {Pendry}, \citenamefont {Chilkoti},\ and\
  \citenamefont {Smith}}]{ciraci2012probing}%
  \BibitemOpen
  \bibfield  {author} {\bibinfo {author} {\bibfnamefont {C.}~\bibnamefont
  {Cirac{\`\i}}}, \bibinfo {author} {\bibfnamefont {R.}~\bibnamefont {Hill}},
  \bibinfo {author} {\bibfnamefont {J.}~\bibnamefont {Mock}}, \bibinfo {author}
  {\bibfnamefont {Y.}~\bibnamefont {Urzhumov}}, \bibinfo {author}
  {\bibfnamefont {A.}~\bibnamefont {Fern{\'a}ndez-Dom{\'\i}nguez}}, \bibinfo
  {author} {\bibfnamefont {S.}~\bibnamefont {Maier}}, \bibinfo {author}
  {\bibfnamefont {J.}~\bibnamefont {Pendry}}, \bibinfo {author} {\bibfnamefont
  {A.}~\bibnamefont {Chilkoti}}, \ and\ \bibinfo {author} {\bibfnamefont
  {D.}~\bibnamefont {Smith}},\ }\href@noop {} {\bibfield  {journal} {\bibinfo
  {journal} {Science}\ }\textbf {\bibinfo {volume} {337}},\ \bibinfo {pages}
  {1072} (\bibinfo {year} {2012})}\BibitemShut {NoStop}%
\bibitem [{\citenamefont {Cirac{\`\i}}\ \emph {et~al.}(2014)\citenamefont
  {Cirac{\`\i}}, \citenamefont {Chen}, \citenamefont {Mock}, \citenamefont
  {McGuire}, \citenamefont {Liu}, \citenamefont {Oh},\ and\ \citenamefont
  {Smith}}]{ciraci2014film}%
  \BibitemOpen
  \bibfield  {author} {\bibinfo {author} {\bibfnamefont {C.}~\bibnamefont
  {Cirac{\`\i}}}, \bibinfo {author} {\bibfnamefont {X.}~\bibnamefont {Chen}},
  \bibinfo {author} {\bibfnamefont {J.~J.}\ \bibnamefont {Mock}}, \bibinfo
  {author} {\bibfnamefont {F.}~\bibnamefont {McGuire}}, \bibinfo {author}
  {\bibfnamefont {X.}~\bibnamefont {Liu}}, \bibinfo {author} {\bibfnamefont
  {S.-H.}\ \bibnamefont {Oh}}, \ and\ \bibinfo {author} {\bibfnamefont {D.~R.}\
  \bibnamefont {Smith}},\ }\href@noop {} {\bibfield  {journal} {\bibinfo
  {journal} {Applied Physics Letters}\ }\textbf {\bibinfo {volume} {104}},\
  \bibinfo {pages} {023109} (\bibinfo {year} {2014})}\BibitemShut {NoStop}%
\bibitem [{\citenamefont {Ciraci}\ \emph {et~al.}(2013)\citenamefont {Ciraci},
  \citenamefont {Pendry},\ and\ \citenamefont
  {Smith}}]{ciraci2013hydrodynamic}%
  \BibitemOpen
  \bibfield  {author} {\bibinfo {author} {\bibfnamefont {C.}~\bibnamefont
  {Ciraci}}, \bibinfo {author} {\bibfnamefont {J.~B.}\ \bibnamefont {Pendry}},
  \ and\ \bibinfo {author} {\bibfnamefont {D.~R.}\ \bibnamefont {Smith}},\
  }\href@noop {} {\bibfield  {journal} {\bibinfo  {journal} {ChemPhysChem}\
  }\textbf {\bibinfo {volume} {14}},\ \bibinfo {pages} {1109} (\bibinfo {year}
  {2013})}\BibitemShut {NoStop}%
\bibitem [{\citenamefont {Scholl}\ \emph {et~al.}(2012)\citenamefont {Scholl},
  \citenamefont {Koh},\ and\ \citenamefont {Dionne}}]{scholl2012quantum}%
  \BibitemOpen
  \bibfield  {author} {\bibinfo {author} {\bibfnamefont {J.~A.}\ \bibnamefont
  {Scholl}}, \bibinfo {author} {\bibfnamefont {A.~L.}\ \bibnamefont {Koh}}, \
  and\ \bibinfo {author} {\bibfnamefont {J.~A.}\ \bibnamefont {Dionne}},\
  }\href@noop {} {\bibfield  {journal} {\bibinfo  {journal} {Nature}\ }\textbf
  {\bibinfo {volume} {483}},\ \bibinfo {pages} {421} (\bibinfo {year}
  {2012})}\BibitemShut {NoStop}%
\bibitem [{\citenamefont {Raza}\ \emph {et~al.}(2013)\citenamefont {Raza},
  \citenamefont {Christensen}, \citenamefont {Wubs}, \citenamefont
  {Bozhevolnyi},\ and\ \citenamefont {Mortensen}}]{raza2013nonlocal}%
  \BibitemOpen
  \bibfield  {author} {\bibinfo {author} {\bibfnamefont {S.}~\bibnamefont
  {Raza}}, \bibinfo {author} {\bibfnamefont {T.}~\bibnamefont {Christensen}},
  \bibinfo {author} {\bibfnamefont {M.}~\bibnamefont {Wubs}}, \bibinfo {author}
  {\bibfnamefont {S.~I.}\ \bibnamefont {Bozhevolnyi}}, \ and\ \bibinfo {author}
  {\bibfnamefont {N.~A.}\ \bibnamefont {Mortensen}},\ }\href@noop {} {\bibfield
   {journal} {\bibinfo  {journal} {Physical Review B}\ }\textbf {\bibinfo
  {volume} {88}},\ \bibinfo {pages} {115401} (\bibinfo {year}
  {2013})}\BibitemShut {NoStop}%
\bibitem [{\citenamefont {Hajisalem}\ \emph {et~al.}(2014)\citenamefont
  {Hajisalem}, \citenamefont {Min}, \citenamefont {Gelfand},\ and\
  \citenamefont {Gordon}}]{hajilsalem14}%
  \BibitemOpen
  \bibfield  {author} {\bibinfo {author} {\bibfnamefont {G.}~\bibnamefont
  {Hajisalem}}, \bibinfo {author} {\bibfnamefont {Q.}~\bibnamefont {Min}},
  \bibinfo {author} {\bibfnamefont {R.}~\bibnamefont {Gelfand}}, \ and\
  \bibinfo {author} {\bibfnamefont {R.}~\bibnamefont {Gordon}},\ }\href
  {\doibase 10.1364/OE.22.009604} {\bibfield  {journal} {\bibinfo  {journal}
  {Opt. Express}\ }\textbf {\bibinfo {volume} {22}},\ \bibinfo {pages} {9604}
  (\bibinfo {year} {2014})}\BibitemShut {NoStop}%
\bibitem [{\citenamefont {Esteban}\ \emph {et~al.}(2012)\citenamefont
  {Esteban}, \citenamefont {Borisov}, \citenamefont {Nordlander},\ and\
  \citenamefont {Aizpurua}}]{esteban2012bridging}%
  \BibitemOpen
  \bibfield  {author} {\bibinfo {author} {\bibfnamefont {R.}~\bibnamefont
  {Esteban}}, \bibinfo {author} {\bibfnamefont {A.~G.}\ \bibnamefont
  {Borisov}}, \bibinfo {author} {\bibfnamefont {P.}~\bibnamefont {Nordlander}},
  \ and\ \bibinfo {author} {\bibfnamefont {J.}~\bibnamefont {Aizpurua}},\
  }\href@noop {} {\bibfield  {journal} {\bibinfo  {journal} {Nature
  communications}\ }\textbf {\bibinfo {volume} {3}},\ \bibinfo {pages} {825}
  (\bibinfo {year} {2012})}\BibitemShut {NoStop}%
\bibitem [{\citenamefont {Savage}\ \emph {et~al.}(2012)\citenamefont {Savage},
  \citenamefont {Hawkeye}, \citenamefont {Esteban}, \citenamefont {Borisov},
  \citenamefont {Aizpurua},\ and\ \citenamefont
  {Baumberg}}]{savage2012revealing}%
  \BibitemOpen
  \bibfield  {author} {\bibinfo {author} {\bibfnamefont {K.~J.}\ \bibnamefont
  {Savage}}, \bibinfo {author} {\bibfnamefont {M.~M.}\ \bibnamefont {Hawkeye}},
  \bibinfo {author} {\bibfnamefont {R.}~\bibnamefont {Esteban}}, \bibinfo
  {author} {\bibfnamefont {A.~G.}\ \bibnamefont {Borisov}}, \bibinfo {author}
  {\bibfnamefont {J.}~\bibnamefont {Aizpurua}}, \ and\ \bibinfo {author}
  {\bibfnamefont {J.~J.}\ \bibnamefont {Baumberg}},\ }\href@noop {} {\bibfield
  {journal} {\bibinfo  {journal} {Nature}\ }\textbf {\bibinfo {volume} {491}},\
  \bibinfo {pages} {574} (\bibinfo {year} {2012})}\BibitemShut {NoStop}%
\bibitem [{\citenamefont {Teperik}\ \emph {et~al.}(2013)\citenamefont
  {Teperik}, \citenamefont {Nordlander}, \citenamefont {Aizpurua},\ and\
  \citenamefont {Borisov}}]{teperik2013robust}%
  \BibitemOpen
  \bibfield  {author} {\bibinfo {author} {\bibfnamefont {T.~V.}\ \bibnamefont
  {Teperik}}, \bibinfo {author} {\bibfnamefont {P.}~\bibnamefont {Nordlander}},
  \bibinfo {author} {\bibfnamefont {J.}~\bibnamefont {Aizpurua}}, \ and\
  \bibinfo {author} {\bibfnamefont {A.~G.}\ \bibnamefont {Borisov}},\
  }\href@noop {} {\bibfield  {journal} {\bibinfo  {journal} {Physical review
  letters}\ }\textbf {\bibinfo {volume} {110}},\ \bibinfo {pages} {263901}
  (\bibinfo {year} {2013})}\BibitemShut {NoStop}%
\bibitem [{\citenamefont {Fern{\'a}ndez-Dom{\'\i}nguez}\ \emph
  {et~al.}(2010)\citenamefont {Fern{\'a}ndez-Dom{\'\i}nguez}, \citenamefont
  {Maier},\ and\ \citenamefont {Pendry}}]{fernandez2010collection}%
  \BibitemOpen
  \bibfield  {author} {\bibinfo {author} {\bibfnamefont {A.}~\bibnamefont
  {Fern{\'a}ndez-Dom{\'\i}nguez}}, \bibinfo {author} {\bibfnamefont
  {S.}~\bibnamefont {Maier}}, \ and\ \bibinfo {author} {\bibfnamefont
  {J.}~\bibnamefont {Pendry}},\ }\href@noop {} {\bibfield  {journal} {\bibinfo
  {journal} {Physical review letters}\ }\textbf {\bibinfo {volume} {105}},\
  \bibinfo {pages} {266807} (\bibinfo {year} {2010})}\BibitemShut {NoStop}%
\bibitem [{\citenamefont {Wiener}\ \emph {et~al.}(2012)\citenamefont {Wiener},
  \citenamefont {Fern{\'a}ndez-Dom{\'\i}nguez}, \citenamefont {Horsfield},
  \citenamefont {Pendry},\ and\ \citenamefont {Maier}}]{wiener2012nonlocal}%
  \BibitemOpen
  \bibfield  {author} {\bibinfo {author} {\bibfnamefont {A.}~\bibnamefont
  {Wiener}}, \bibinfo {author} {\bibfnamefont {A.~I.}\ \bibnamefont
  {Fern{\'a}ndez-Dom{\'\i}nguez}}, \bibinfo {author} {\bibfnamefont {A.~P.}\
  \bibnamefont {Horsfield}}, \bibinfo {author} {\bibfnamefont {J.~B.}\
  \bibnamefont {Pendry}}, \ and\ \bibinfo {author} {\bibfnamefont {S.~A.}\
  \bibnamefont {Maier}},\ }\href@noop {} {\bibfield  {journal} {\bibinfo
  {journal} {Nano letters}\ }\textbf {\bibinfo {volume} {12}},\ \bibinfo
  {pages} {3308} (\bibinfo {year} {2012})}\BibitemShut {NoStop}%
\bibitem [{\citenamefont {Moreau}\ \emph {et~al.}(2013)\citenamefont {Moreau},
  \citenamefont {Cirac{\`\i}},\ and\ \citenamefont {Smith}}]{moreau13}%
  \BibitemOpen
  \bibfield  {author} {\bibinfo {author} {\bibfnamefont {A.}~\bibnamefont
  {Moreau}}, \bibinfo {author} {\bibfnamefont {C.}~\bibnamefont {Cirac{\`\i}}},
  \ and\ \bibinfo {author} {\bibfnamefont {D.~R.}\ \bibnamefont {Smith}},\
  }\href@noop {} {\bibfield  {journal} {\bibinfo  {journal} {Physical Review
  B}\ }\textbf {\bibinfo {volume} {87}},\ \bibinfo {pages} {045401} (\bibinfo
  {year} {2013})}\BibitemShut {NoStop}%
\bibitem [{\citenamefont {Dechaux}\ \emph {et~al.}(2016)\citenamefont
  {Dechaux}, \citenamefont {Tichit}, \citenamefont {Cirac{\`\i}}, \citenamefont
  {Benedicto}, \citenamefont {Poll{\`e}s}, \citenamefont {Centeno},
  \citenamefont {Smith},\ and\ \citenamefont {Moreau}}]{dechaux16}%
  \BibitemOpen
  \bibfield  {author} {\bibinfo {author} {\bibfnamefont {M.}~\bibnamefont
  {Dechaux}}, \bibinfo {author} {\bibfnamefont {P.-H.}\ \bibnamefont {Tichit}},
  \bibinfo {author} {\bibfnamefont {C.}~\bibnamefont {Cirac{\`\i}}}, \bibinfo
  {author} {\bibfnamefont {J.}~\bibnamefont {Benedicto}}, \bibinfo {author}
  {\bibfnamefont {R.}~\bibnamefont {Poll{\`e}s}}, \bibinfo {author}
  {\bibfnamefont {E.}~\bibnamefont {Centeno}}, \bibinfo {author} {\bibfnamefont
  {D.~R.}\ \bibnamefont {Smith}}, \ and\ \bibinfo {author} {\bibfnamefont
  {A.}~\bibnamefont {Moreau}},\ }\href@noop {} {\bibfield  {journal} {\bibinfo
  {journal} {Physical Review B}\ }\textbf {\bibinfo {volume} {93}},\ \bibinfo
  {pages} {045413} (\bibinfo {year} {2016})}\BibitemShut {NoStop}%
\bibitem [{\citenamefont {Toscano}\ \emph {et~al.}(2012)\citenamefont
  {Toscano}, \citenamefont {Raza}, \citenamefont {Jauho}, \citenamefont
  {Mortensen},\ and\ \citenamefont {Wubs}}]{toscano2012modified}%
  \BibitemOpen
  \bibfield  {author} {\bibinfo {author} {\bibfnamefont {G.}~\bibnamefont
  {Toscano}}, \bibinfo {author} {\bibfnamefont {S.}~\bibnamefont {Raza}},
  \bibinfo {author} {\bibfnamefont {A.-P.}\ \bibnamefont {Jauho}}, \bibinfo
  {author} {\bibfnamefont {N.~A.}\ \bibnamefont {Mortensen}}, \ and\ \bibinfo
  {author} {\bibfnamefont {M.}~\bibnamefont {Wubs}},\ }\href@noop {} {\bibfield
   {journal} {\bibinfo  {journal} {Optics express}\ }\textbf {\bibinfo {volume}
  {20}},\ \bibinfo {pages} {4176} (\bibinfo {year} {2012})}\BibitemShut
  {NoStop}%
\bibitem [{\citenamefont {Benedicto}\ \emph {et~al.}(2015)\citenamefont
  {Benedicto}, \citenamefont {Poll{\`e}s}, \citenamefont {Cirac{\`\i}},
  \citenamefont {Centeno}, \citenamefont {Smith},\ and\ \citenamefont
  {Moreau}}]{benedicto2015numerical}%
  \BibitemOpen
  \bibfield  {author} {\bibinfo {author} {\bibfnamefont {J.}~\bibnamefont
  {Benedicto}}, \bibinfo {author} {\bibfnamefont {R.}~\bibnamefont
  {Poll{\`e}s}}, \bibinfo {author} {\bibfnamefont {C.}~\bibnamefont
  {Cirac{\`\i}}}, \bibinfo {author} {\bibfnamefont {E.}~\bibnamefont
  {Centeno}}, \bibinfo {author} {\bibfnamefont {D.~R.}\ \bibnamefont {Smith}},
  \ and\ \bibinfo {author} {\bibfnamefont {A.}~\bibnamefont {Moreau}},\
  }\href@noop {} {\bibfield  {journal} {\bibinfo  {journal} {JOSA A}\ }\textbf
  {\bibinfo {volume} {32}},\ \bibinfo {pages} {1581} (\bibinfo {year}
  {2015})}\BibitemShut {NoStop}%
\bibitem [{\citenamefont {Schmitt}\ \emph {et~al.}(2016)\citenamefont
  {Schmitt}, \citenamefont {Scheid}, \citenamefont {Lanteri}, \citenamefont
  {Moreau},\ and\ \citenamefont {Viquerat}}]{Schmitt2016396}%
  \BibitemOpen
  \bibfield  {author} {\bibinfo {author} {\bibfnamefont {N.}~\bibnamefont
  {Schmitt}}, \bibinfo {author} {\bibfnamefont {C.}~\bibnamefont {Scheid}},
  \bibinfo {author} {\bibfnamefont {S.}~\bibnamefont {Lanteri}}, \bibinfo
  {author} {\bibfnamefont {A.}~\bibnamefont {Moreau}}, \ and\ \bibinfo {author}
  {\bibfnamefont {J.}~\bibnamefont {Viquerat}},\ }\href {\doibase
  http://dx.doi.org/10.1016/j.jcp.2016.04.020} {\bibfield  {journal} {\bibinfo
  {journal} {Journal of Computational Physics}\ }\textbf {\bibinfo {volume}
  {316}},\ \bibinfo {pages} {396 } (\bibinfo {year} {2016})}\BibitemShut
  {NoStop}%
\bibitem [{\citenamefont {Haberland}(2013)}]{haberland2013looking}%
  \BibitemOpen
  \bibfield  {author} {\bibinfo {author} {\bibfnamefont {H.}~\bibnamefont
  {Haberland}},\ }\href@noop {} {\bibfield  {journal} {\bibinfo  {journal}
  {Nature}\ }\textbf {\bibinfo {volume} {494}},\ \bibinfo {pages} {E1}
  (\bibinfo {year} {2013})}\BibitemShut {NoStop}%
\bibitem [{\citenamefont {Toscano}\ \emph {et~al.}(2015)\citenamefont
  {Toscano}, \citenamefont {Straubel}, \citenamefont {Kwiatkowski},
  \citenamefont {Rockstuhl}, \citenamefont {Evers}, \citenamefont {Xu},
  \citenamefont {Mortensen},\ and\ \citenamefont
  {Wubs}}]{toscano2015resonance}%
  \BibitemOpen
  \bibfield  {author} {\bibinfo {author} {\bibfnamefont {G.}~\bibnamefont
  {Toscano}}, \bibinfo {author} {\bibfnamefont {J.}~\bibnamefont {Straubel}},
  \bibinfo {author} {\bibfnamefont {A.}~\bibnamefont {Kwiatkowski}}, \bibinfo
  {author} {\bibfnamefont {C.}~\bibnamefont {Rockstuhl}}, \bibinfo {author}
  {\bibfnamefont {F.}~\bibnamefont {Evers}}, \bibinfo {author} {\bibfnamefont
  {H.}~\bibnamefont {Xu}}, \bibinfo {author} {\bibfnamefont {N.~A.}\
  \bibnamefont {Mortensen}}, \ and\ \bibinfo {author} {\bibfnamefont
  {M.}~\bibnamefont {Wubs}},\ }\href@noop {} {\bibfield  {journal} {\bibinfo
  {journal} {Nature communications}\ }\textbf {\bibinfo {volume} {6}},\
  \bibinfo {pages} {7132} (\bibinfo {year} {2015})}\BibitemShut {NoStop}%
\bibitem [{\citenamefont {Ciraci}\ and\ \citenamefont
  {Della~Sala}(2016)}]{ciraci2016quantum}%
  \BibitemOpen
  \bibfield  {author} {\bibinfo {author} {\bibfnamefont {C.}~\bibnamefont
  {Ciraci}}\ and\ \bibinfo {author} {\bibfnamefont {F.}~\bibnamefont
  {Della~Sala}},\ }\href@noop {} {\bibfield  {journal} {\bibinfo  {journal}
  {Physical Review B}\ }\textbf {\bibinfo {volume} {93}},\ \bibinfo {pages}
  {205405} (\bibinfo {year} {2016})}\BibitemShut {NoStop}%
\bibitem [{\citenamefont {{Lopez-Rios, T.}}\ \emph {et~al.}(1979)\citenamefont
  {{Lopez-Rios, T.}}, \citenamefont {{Abel\`es, F.}},\ and\ \citenamefont
  {{Vuye, G.}}}]{lopezrios}%
  \BibitemOpen
  \bibfield  {author} {\bibinfo {author} {\bibnamefont {{Lopez-Rios, T.}}},
  \bibinfo {author} {\bibnamefont {{Abel\`es, F.}}}, \ and\ \bibinfo {author}
  {\bibnamefont {{Vuye, G.}}},\ }\href {\doibase
  10.1051/jphyslet:019790040014034300} {\bibfield  {journal} {\bibinfo
  {journal} {J. Physique Lett.}\ }\textbf {\bibinfo {volume} {40}},\ \bibinfo
  {pages} {343} (\bibinfo {year} {1979})}\BibitemShut {NoStop}%
\bibitem [{\citenamefont {Bozhevolnyi}\ and\ \citenamefont
  {S{\o}ndergaard}(2007)}]{bozhevolnyi2007general}%
  \BibitemOpen
  \bibfield  {author} {\bibinfo {author} {\bibfnamefont {S.~I.}\ \bibnamefont
  {Bozhevolnyi}}\ and\ \bibinfo {author} {\bibfnamefont {T.}~\bibnamefont
  {S{\o}ndergaard}},\ }\href@noop {} {\bibfield  {journal} {\bibinfo  {journal}
  {Optics express}\ }\textbf {\bibinfo {volume} {15}},\ \bibinfo {pages}
  {10869} (\bibinfo {year} {2007})}\BibitemShut {NoStop}%
\bibitem [{\citenamefont {Berini}(2009)}]{berini2009long}%
  \BibitemOpen
  \bibfield  {author} {\bibinfo {author} {\bibfnamefont {P.}~\bibnamefont
  {Berini}},\ }\href@noop {} {\bibfield  {journal} {\bibinfo  {journal}
  {Advances in optics and photonics}\ }\textbf {\bibinfo {volume} {1}},\
  \bibinfo {pages} {484} (\bibinfo {year} {2009})}\BibitemShut {NoStop}%
\bibitem [{\citenamefont {Berini}(2000)}]{PhysRevB.61.10484}%
  \BibitemOpen
  \bibfield  {author} {\bibinfo {author} {\bibfnamefont {P.}~\bibnamefont
  {Berini}},\ }\href {\doibase 10.1103/PhysRevB.61.10484} {\bibfield  {journal}
  {\bibinfo  {journal} {Phys. Rev. B}\ }\textbf {\bibinfo {volume} {61}},\
  \bibinfo {pages} {10484} (\bibinfo {year} {2000})}\BibitemShut {NoStop}%
\bibitem [{\citenamefont {Pitelet}\ \emph {et~al.}(2017)\citenamefont
  {Pitelet}, \citenamefont {Mallet}, \citenamefont {Centeno},\ and\
  \citenamefont {Moreau}}]{pitelet17}%
  \BibitemOpen
  \bibfield  {author} {\bibinfo {author} {\bibfnamefont {A.}~\bibnamefont
  {Pitelet}}, \bibinfo {author} {\bibfnamefont {E.}~\bibnamefont {Mallet}},
  \bibinfo {author} {\bibfnamefont {E.}~\bibnamefont {Centeno}}, \ and\
  \bibinfo {author} {\bibfnamefont {A.}~\bibnamefont {Moreau}},\ }\href
  {\doibase 10.1103/PhysRevB.96.041406} {\bibfield  {journal} {\bibinfo
  {journal} {Phys. Rev. B}\ }\textbf {\bibinfo {volume} {96}},\ \bibinfo
  {pages} {041406} (\bibinfo {year} {2017})}\BibitemShut {NoStop}%
\bibitem [{\citenamefont {Rakic}\ \emph {et~al.}(1998)\citenamefont {Rakic},
  \citenamefont {Djuri{\v{s}}ic}, \citenamefont {Elazar},\ and\ \citenamefont
  {Majewski}}]{rakic98}%
  \BibitemOpen
  \bibfield  {author} {\bibinfo {author} {\bibfnamefont {A.~D.}\ \bibnamefont
  {Rakic}}, \bibinfo {author} {\bibfnamefont {A.~B.}\ \bibnamefont
  {Djuri{\v{s}}ic}}, \bibinfo {author} {\bibfnamefont {J.~M.}\ \bibnamefont
  {Elazar}}, \ and\ \bibinfo {author} {\bibfnamefont {M.~L.}\ \bibnamefont
  {Majewski}},\ }\href@noop {} {\bibfield  {journal} {\bibinfo  {journal}
  {Applied Optics}\ }\textbf {\bibinfo {volume} {37}},\ \bibinfo {pages} {5271}
  (\bibinfo {year} {1998})}\BibitemShut {NoStop}%
\bibitem [{\citenamefont {Defrance}\ \emph {et~al.}(2016)\citenamefont
  {Defrance}, \citenamefont {Lema{\^\i}tre}, \citenamefont {Ajib},
  \citenamefont {Benedicto}, \citenamefont {Mallet}, \citenamefont
  {Poll{\`e}s}, \citenamefont {Plumey}, \citenamefont {Mihailovic},
  \citenamefont {Centeno}, \citenamefont {Cirac{\`\i}}, \citenamefont {Smith},\
  and\ \citenamefont {Moreau}}]{defrance16}%
  \BibitemOpen
  \bibfield  {author} {\bibinfo {author} {\bibfnamefont {J.}~\bibnamefont
  {Defrance}}, \bibinfo {author} {\bibfnamefont {C.}~\bibnamefont
  {Lema{\^\i}tre}}, \bibinfo {author} {\bibfnamefont {R.}~\bibnamefont {Ajib}},
  \bibinfo {author} {\bibfnamefont {J.}~\bibnamefont {Benedicto}}, \bibinfo
  {author} {\bibfnamefont {E.}~\bibnamefont {Mallet}}, \bibinfo {author}
  {\bibfnamefont {R.}~\bibnamefont {Poll{\`e}s}}, \bibinfo {author}
  {\bibfnamefont {J.-P.}\ \bibnamefont {Plumey}}, \bibinfo {author}
  {\bibfnamefont {M.}~\bibnamefont {Mihailovic}}, \bibinfo {author}
  {\bibfnamefont {E.}~\bibnamefont {Centeno}}, \bibinfo {author} {\bibfnamefont
  {C.}~\bibnamefont {Cirac{\`\i}}}, \bibinfo {author} {\bibfnamefont
  {D.}~\bibnamefont {Smith}}, \ and\ \bibinfo {author} {\bibfnamefont
  {A.}~\bibnamefont {Moreau}},\ }\href@noop {} {\bibfield  {journal} {\bibinfo
  {journal} {Journal of Open Research Software}\ }\textbf {\bibinfo {volume}
  {4}} (\bibinfo {year} {2016})}\BibitemShut {NoStop}%
\bibitem [{\citenamefont {DeVore}(1951)}]{DeVore:51}%
  \BibitemOpen
  \bibfield  {author} {\bibinfo {author} {\bibfnamefont {J.~R.}\ \bibnamefont
  {DeVore}},\ }\href {\doibase 10.1364/JOSA.41.000416} {\bibfield  {journal}
  {\bibinfo  {journal} {J. Opt. Soc. Am.}\ }\textbf {\bibinfo {volume} {41}},\
  \bibinfo {pages} {416} (\bibinfo {year} {1951})}\BibitemShut {NoStop}%
\bibitem [{\citenamefont {Ritchie}\ and\ \citenamefont
  {Eldridge}(1962)}]{ritchie1962optical}%
  \BibitemOpen
  \bibfield  {author} {\bibinfo {author} {\bibfnamefont {R.}~\bibnamefont
  {Ritchie}}\ and\ \bibinfo {author} {\bibfnamefont {H.}~\bibnamefont
  {Eldridge}},\ }\href@noop {} {\bibfield  {journal} {\bibinfo  {journal}
  {Physical Review}\ }\textbf {\bibinfo {volume} {126}},\ \bibinfo {pages}
  {1935} (\bibinfo {year} {1962})}\BibitemShut {NoStop}%
\bibitem [{\citenamefont {Raza}\ \emph {et~al.}(2015)\citenamefont {Raza},
  \citenamefont {Bozhevolnyi}, \citenamefont {Wubs},\ and\ \citenamefont
  {Mortensen}}]{raza2015nonlocal}%
  \BibitemOpen
  \bibfield  {author} {\bibinfo {author} {\bibfnamefont {S.}~\bibnamefont
  {Raza}}, \bibinfo {author} {\bibfnamefont {S.~I.}\ \bibnamefont
  {Bozhevolnyi}}, \bibinfo {author} {\bibfnamefont {M.}~\bibnamefont {Wubs}}, \
  and\ \bibinfo {author} {\bibfnamefont {N.~A.}\ \bibnamefont {Mortensen}},\
  }\href@noop {} {\bibfield  {journal} {\bibinfo  {journal} {Journal of
  Physics: Condensed Matter}\ }\textbf {\bibinfo {volume} {27}},\ \bibinfo
  {pages} {183204} (\bibinfo {year} {2015})}\BibitemShut {NoStop}%
\bibitem [{\citenamefont {Akselrod}\ \emph {et~al.}(2014)\citenamefont
  {Akselrod}, \citenamefont {Argyropoulos}, \citenamefont {Hoang},
  \citenamefont {Cirac{\`\i}}, \citenamefont {Fang}, \citenamefont {Huang},
  \citenamefont {Smith},\ and\ \citenamefont
  {Mikkelsen}}]{akselrod2014probing}%
  \BibitemOpen
  \bibfield  {author} {\bibinfo {author} {\bibfnamefont {G.~M.}\ \bibnamefont
  {Akselrod}}, \bibinfo {author} {\bibfnamefont {C.}~\bibnamefont
  {Argyropoulos}}, \bibinfo {author} {\bibfnamefont {T.~B.}\ \bibnamefont
  {Hoang}}, \bibinfo {author} {\bibfnamefont {C.}~\bibnamefont {Cirac{\`\i}}},
  \bibinfo {author} {\bibfnamefont {C.}~\bibnamefont {Fang}}, \bibinfo {author}
  {\bibfnamefont {J.}~\bibnamefont {Huang}}, \bibinfo {author} {\bibfnamefont
  {D.~R.}\ \bibnamefont {Smith}}, \ and\ \bibinfo {author} {\bibfnamefont
  {M.~H.}\ \bibnamefont {Mikkelsen}},\ }\href@noop {} {\bibfield  {journal}
  {\bibinfo  {journal} {Nature Photonics}\ }\textbf {\bibinfo {volume} {8}},\
  \bibinfo {pages} {835} (\bibinfo {year} {2014})}\BibitemShut {NoStop}%
\bibitem [{\citenamefont {Lassiter}\ \emph {et~al.}(2014)\citenamefont
  {Lassiter}, \citenamefont {Chen}, \citenamefont {Liu}, \citenamefont
  {Cirac{\`\i}}, \citenamefont {Hoang}, \citenamefont {Larouche}, \citenamefont
  {Oh}, \citenamefont {Mikkelsen},\ and\ \citenamefont
  {Smith}}]{lassiter2014third}%
  \BibitemOpen
  \bibfield  {author} {\bibinfo {author} {\bibfnamefont {J.~B.}\ \bibnamefont
  {Lassiter}}, \bibinfo {author} {\bibfnamefont {X.}~\bibnamefont {Chen}},
  \bibinfo {author} {\bibfnamefont {X.}~\bibnamefont {Liu}}, \bibinfo {author}
  {\bibfnamefont {C.}~\bibnamefont {Cirac{\`\i}}}, \bibinfo {author}
  {\bibfnamefont {T.~B.}\ \bibnamefont {Hoang}}, \bibinfo {author}
  {\bibfnamefont {S.}~\bibnamefont {Larouche}}, \bibinfo {author}
  {\bibfnamefont {S.-H.}\ \bibnamefont {Oh}}, \bibinfo {author} {\bibfnamefont
  {M.~H.}\ \bibnamefont {Mikkelsen}}, \ and\ \bibinfo {author} {\bibfnamefont
  {D.~R.}\ \bibnamefont {Smith}},\ }\href@noop {} {\bibfield  {journal}
  {\bibinfo  {journal} {Acs Photonics}\ }\textbf {\bibinfo {volume} {1}},\
  \bibinfo {pages} {1212} (\bibinfo {year} {2014})}\BibitemShut {NoStop}%
\bibitem [{\citenamefont {Haffner}\ \emph {et~al.}(2015)\citenamefont
  {Haffner}, \citenamefont {Heni}, \citenamefont {Fedoryshyn}, \citenamefont
  {Niegemann}, \citenamefont {Melikyan}, \citenamefont {Elder}, \citenamefont
  {Baeuerle}, \citenamefont {Salamin}, \citenamefont {Josten}, \citenamefont
  {Koch}, \citenamefont {Hoessbacher}, \citenamefont {Ducry}, \citenamefont
  {Juchli}, \citenamefont {Emboras}, \citenamefont {Hillerkuss}, \citenamefont
  {Kohl}, \citenamefont {Dalton}, \citenamefont {Hafner},\ and\ \citenamefont
  {Leuthold}}]{haffner2015all}%
  \BibitemOpen
  \bibfield  {author} {\bibinfo {author} {\bibfnamefont {C.}~\bibnamefont
  {Haffner}}, \bibinfo {author} {\bibfnamefont {W.}~\bibnamefont {Heni}},
  \bibinfo {author} {\bibfnamefont {Y.}~\bibnamefont {Fedoryshyn}}, \bibinfo
  {author} {\bibfnamefont {J.}~\bibnamefont {Niegemann}}, \bibinfo {author}
  {\bibfnamefont {A.}~\bibnamefont {Melikyan}}, \bibinfo {author}
  {\bibfnamefont {D.~L.}\ \bibnamefont {Elder}}, \bibinfo {author}
  {\bibfnamefont {B.}~\bibnamefont {Baeuerle}}, \bibinfo {author}
  {\bibfnamefont {Y.}~\bibnamefont {Salamin}}, \bibinfo {author} {\bibfnamefont
  {A.}~\bibnamefont {Josten}}, \bibinfo {author} {\bibfnamefont
  {U.}~\bibnamefont {Koch}}, \bibinfo {author} {\bibfnamefont {C.}~\bibnamefont
  {Hoessbacher}}, \bibinfo {author} {\bibfnamefont {F.}~\bibnamefont {Ducry}},
  \bibinfo {author} {\bibfnamefont {L.}~\bibnamefont {Juchli}}, \bibinfo
  {author} {\bibfnamefont {A.}~\bibnamefont {Emboras}}, \bibinfo {author}
  {\bibfnamefont {D.}~\bibnamefont {Hillerkuss}}, \bibinfo {author}
  {\bibfnamefont {M.}~\bibnamefont {Kohl}}, \bibinfo {author} {\bibfnamefont
  {L.~R.}\ \bibnamefont {Dalton}}, \bibinfo {author} {\bibfnamefont
  {C.}~\bibnamefont {Hafner}}, \ and\ \bibinfo {author} {\bibfnamefont
  {J.}~\bibnamefont {Leuthold}},\ }\href@noop {} {\bibfield  {journal}
  {\bibinfo  {journal} {Nature Photonics}\ }\textbf {\bibinfo {volume} {9}},\
  \bibinfo {pages} {525} (\bibinfo {year} {2015})}\BibitemShut {NoStop}%
\bibitem [{\citenamefont {Nielsen}\ \emph {et~al.}(2017)\citenamefont
  {Nielsen}, \citenamefont {Shi}, \citenamefont {Dichtl}, \citenamefont
  {Maier},\ and\ \citenamefont {Oulton}}]{nielsen2017giant}%
  \BibitemOpen
  \bibfield  {author} {\bibinfo {author} {\bibfnamefont {M.~P.}\ \bibnamefont
  {Nielsen}}, \bibinfo {author} {\bibfnamefont {X.}~\bibnamefont {Shi}},
  \bibinfo {author} {\bibfnamefont {P.}~\bibnamefont {Dichtl}}, \bibinfo
  {author} {\bibfnamefont {S.~A.}\ \bibnamefont {Maier}}, \ and\ \bibinfo
  {author} {\bibfnamefont {R.~F.}\ \bibnamefont {Oulton}},\ }\href@noop {}
  {\bibfield  {journal} {\bibinfo  {journal} {Science}\ }\textbf {\bibinfo
  {volume} {358}},\ \bibinfo {pages} {1179} (\bibinfo {year}
  {2017})}\BibitemShut {NoStop}%
\end{thebibliography}%
 
\end{document}